\documentclass[11pt,reqno]{amsart}
\usepackage{t1enc}
\usepackage[latin1]{inputenc}
\usepackage{amsmath,latexsym,amssymb,graphicx,dsfont,amsthm,amsfonts}
\usepackage{color}
\usepackage{umoline}
\usepackage{float}
\usepackage{apalike}

\renewcommand{\O}{\mathbf{O}}

\newcommand{\elo}[1]{\mathrm{Elo}_{#1}}

\begin{document}

\numberwithin{equation}{section}

\title[Prediction models for the FIFA World Cup 2018]{On Elo based prediction models for the FIFA Worldcup 2018}
\author{Lorenz A. Gilch and Sebastian M\"uller}


\address{Lorenz Gilch: Universit\"at Passau, Innstrasse 33, 94032 Passau, Germany}
\address{Sebastian M\"uller: Aix-Marseille Universit\'e, 39, rue F. Joliot Curie,
13453 Marseille Cedex 13, France}

\email{gilch@TUGraz.at, dr.sebastian.mueller@gmail.com}
\urladdr{http://www.math.tugraz.at/$\sim$gilch/, https://sites.google.com/site/drsebastianmueller/}
\date{\today}

\maketitle

\begin{abstract}
We propose  an approach for the analysis and prediction of a football championship. It is based on  Poisson regression models that include the Elo points of the teams as covariates and incorporates differences of team-specific effects. These models for the prediction of the FIFA World Cup 2018 are fitted on all football games on neutral ground of the participating teams since 2010. Based on the model estimates for single matches Monte-Carlo simulations are used to estimate probabilities for reaching the different stages in the FIFA World Cup 2018 for all teams.  

We propose two score functions for ordinal random variables that serve together with the rank probability score  for the  validation of our models with the results of the FIFA World Cups 2010 and 2014.  All  models favor Germany as the new FIFA World Champion. 

All possible  courses of the tournament and their probabilities  are visualized using a single Sankey diagram. 
\end{abstract}

\section{Introduction}
Football is a typical low-scoring game and games are frequently decided through single events in the game. These events may be extraordinary individual performances,  individual errors, injuries,  refereeing errors or just lucky coincidences.  Moreover, during a tournament there are most of the time teams and players that are in exceptional shape and have a strong influence on the outcome of the tournament. One consequence is that every now and then alleged  underdogs win tournaments and reputed favorites drop out already in the group phase.

The above  effects are notoriously difficult to forecast.
Despite this fact, every team has its strengths and weaknesses (e.g. defense and attack) and most of the results reflect the qualities of the teams.  In order to model the random effects and the ``deterministic'' drift  forecasts should be given in terms of  probabilities.

Among football experts and fans alike there is mostly a consensus on the top favorites, e.g. Brazil, Germany, Spain, and more debate on possible underdogs. However, most of these predictions rely on subjective opinions and are not quantifiable. An additional difficulty is the complexity of the tournament, with billions of different outcomes, making it very difficult to obtain accurate guesses of the probabilities of certain events.


A series of statistical models have therefore been proposed in the literature for the prediction of football outcomes. They can be divided into two broad categories. The  first one, the result-based model, models  directly the probability of a game outcome (win/loss/draw), while the second one, the score-based model, focusses on the match score. We use the second approach since the match score is important in the group phase of the championship and it also implies a model for the first one. 

 There are several models for this purpose and most of them involve a Poisson model.  The easiest model,  \cite{Le:97}, assumes independence of the goals scored by each team and that each score can be modeled by a Poisson regression model.  Bivariate Poisson models  were proposed earlier by \cite{Ma:82} and extended by  \cite{DiCo:97} and  \cite{KaNt:03}. A short overview on different Poisson models  and related models like generalised Poisson models or zero-inflated models are given in \cite{ZeKlJa:08} and  \cite{ChSt:11}. 
 Possible covariates for the above models may be divided into two major categories: those  containing ``prospective'' informations and those containing ``retrospective'' informations. The first category contains other forecasts, especially bookmakers' odds, see e.g.  \cite{LeZeHo:10a}, \cite{LeZeHo:12} and references therein. This approach relies on the fact that bookmakers have a strong economic incentive to rate the result correctly and that they can be seen as experts in the matter of the forecast of sport events. However,  their forecast models remain undisclosed and rely on  information that is not publicly available.  
The second category contains only historical data and no other forecasts.
Since models based on the second category allow to explicitly model the influence of the covariates, we pursue this approach using regression models for the outcome of single matches.  
Since the  FIFA World Cup 2018 is a more complex tournament, involving for instance effects such as group draws, e.g. see  \cite{De:11}, and dependences of the different matches,   we  use Monte-Carlo simulations to forecast the whole course of the tournament. For a more detailed summary on statistical modeling of major international football events we refer to \cite{GrScTu:15}  and references therein. 

These days a lot of data on possible  covariates for the forecast models is available.  \cite{GrScTu:15} performed a variable selection on various covariates and found that the three most significant retrospective covariates are the FIFA ranking followed by the number of Champions league and Euro league players of a team. We prefer to consider the Elo ranking instead of the FIFA ranking, since the calculation of the FIFA ranking changed over time and the Elo ranking is  more widely used in football forecast models. See also \cite{GaRo:16} for a recent discussion on this topic and a justification of the Elo ranking. At the time of our analysis the composition and the line ups of the teams have not been announced and hence the two other covariates are not available.
This is  one of the reasons that our models are solely based  on the Elo points and matches  of the participating teams on neutral ground since 2010.   Our results show that, despite the simplicity of the models, the forecasts are conclusive and give together with the visualization, see Figure \ref{fig:sankey}, a concise idea of the possible courses of the tournament.

We propose four models of Poisson regressions with increasing complexity. The validation of the models involve goodness of fit tests and  analysis of residuals and AIC. Moreover, we validate the  models on the FIFA Worldcups 2010 and 2014. This turned out to be a challenging task and the approach we propose here can only be considered as a first step. A first difficulty is that every outcome of each single match is modeled as $G_{A}$:$G_{B}$, where $G_{A}$ (resp.~$G_{B}$) is the number of goals of team A (resp.~of team B). To our knowledge there is no established score function for such kind of pairs of random variables. Even for the easier game outcome (win/loss/draw) there seems not to be a well established candidate for a good score function. However,  the ranked probability skill score (RPS) is a natural and promising candidate; we refer to  \cite{CoFe:12} for a discussion on this topic.  So much the worse we forecast not only a single match but the course of the whole tournament. Even the most probable tournament outcome has a probability, very close to zero  to be actually realized. Hence, deviations of the true tournament outcome from the model's most probable one are not only possible, but most likely. However, simulations of the tournament yield estimates of the  probabilities for each team for reaching the different stages of the tournament. In this way we obtain for each team an ordinal random variable. For this variable we  propose two new score functions and compare them to the RPS and the Brier score. 

The models show a good fit and the score function on the validation on the FIFA Worldcups  2010 and 2014 are very close to each other. This may be surprising since the actual probabilities that a given team wins the cup may be significantly different. However, all models favor Germany (followed by Brazil) to win the FIFA Worldcup 2018.

\section{The models}
Our models are based on the World Football Elo ratings of the teams. It is based on the Elo rating system, see  \cite{Elo:78}, but includes modifications to take various football-specific variables into account. The Elo ranking is published by the website \texttt{eloratings.net}.
 The Elo ratings as they were  on 28 march 2018  for the top $5$ nations (in this rating) are as follows:
\begin{center}
\begin{tabular}{c|c|c|c|c}
\hline
Brazil & Germany & Spain & Argentina & France \cr
\hline
2131 & 	2092    & 	2048	 & 1985	& 1984 \cr
\hline
\end{tabular}
\end{center}
In the next sections we present several models in {\it increasing} complexity. 
They forecast  the outcome of a match between teams $A$ and $B$ as $G_A$ : $G_B$, where $G_A$ (resp. $G_{B}$) is the number of goals scored by team $A$ (resp. $B$).
The  models  are based on Poisson regression models. In these models we assume  $(G_{A}, G_{B})$ to be a bivariate Poisson distributed random variable;
see  Section \ref{sec:discussion} for a discussion on other underlying distributions for $G_A$ and $G_B$. The distribution of $(G_{A}, G_{B})$  
will depend on $A$ and $B$, and the Elo rankings $\elo{A}$ and $\elo{B}$ of the two teams. The models are fitted  
using  all matches of FIFA World Cup 2018 participating teams on \textit{neutral} playground between 1.1.2010 and 31.12.2017; see Section \ref{sec:discussion} for a discussion why we dropped other games.

All  models are key ingredients in order to simulate the whole tournament and to determine the likelihood of the success for each participant.

\subsection{Independent Poisson regression model}
\label{subsec:independent-PR} 
\nopagebreak
In this model we assume both $G_A$ and $G_B$ to be independent Poisson distributed variables with rates $\lambda_{A|B}$ and $\lambda_{B|A}$. We estimate the Poisson rates $\lambda_{A|B}$ and $\lambda_{B|A}$ via Poisson regression with  the Elo scores of $A$ and $B$ as covariates. Poisson regression models are performed for every team in order to incorporate team specific strengths (attack and defense).  The rates are calculated as follows:
\begin{enumerate}
\item The first step models the number of goals $\tilde G_{A}$ scored  by team $A$ playing against a team with a given Elo score $\elo{}=\elo{B}$. The random variable $\tilde G_{A}$  is modeled as a Poisson distribution with parameter $\mu_{A}$. The parameter $\mu_{A}$ as a function of the Elo rating $\elo{\O}$ of the opponent $\O$ is given as
\begin{equation}\label{equ:independent-regression1}
\log \mu_A(\elo{\O}) = \alpha_0 + \alpha_1 \cdot \elo{\O},
\end{equation}
where $\alpha_0$ and $\alpha_1$ are obtained via Poisson regression.  
\item Teams of similar Elo scores  may have different strengths in attack and defense. To take this effect into account  we model the  number of goals team $B$ receives  against a team of Elo score  $\elo{}=\elo{A}$ using a Poisson distribution with parameter $\nu_{B}$. The parameter $\nu_{B}$ as a function of the Elo rating $\elo{\O}$ is given as
\begin{equation}\label{equ:independent-regression2}
\log \nu_B(\elo{\O}) = \beta_0 + \beta_1 \cdot \elo{\O},
\end{equation}
where the parameters $\beta_0$ and $\beta_1$ are obtained via Poisson regression.
\item Team $A$  shall in average score $\mu_A\bigr(\elo{B}\bigr)$ goals against team $B$, but team $B$ shall have $\nu_B\bigl(\elo{A}\bigr)$ goals against. As these two values rarely coincides we model the numbers of goals $G_A$ as a Poisson distribution with parameter
$$
\lambda = \lambda_{A|B} = \frac{\mu_A\bigl(\elo{B}\bigr)+\nu_B\bigl(\elo{A}\bigr)}{2}.
$$
Analogously, we obtain
$$
\lambda_{B|A} = \frac{\mu_B\bigl(\elo{A}\bigr)+\nu_A\bigl(\elo{B}\bigr)}{2}.
$$
\end{enumerate}
For each team, the regression parameters $\alpha_0, \alpha_1, \beta_0$ and $\beta_1$ are estimated. The match $A$ versus $B$ is then  simulated using two independent Poisson random variables $G_A$ and $G_B$ with rates $\lambda_{A|B}$ and $\lambda_{B|A}$.

\subsubsection{Regression plots}

\begin{figure}[ht]
\begin{center}
\includegraphics[width=6cm]{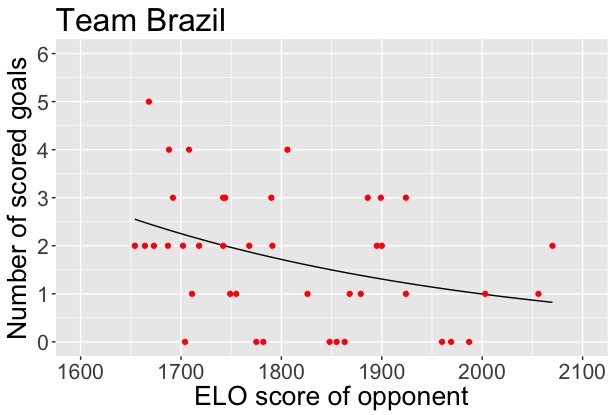}
\hfill
\includegraphics[width=6cm]{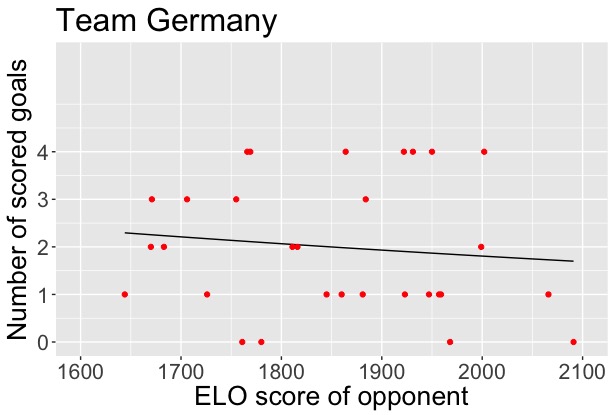} 
\end{center}
\caption{Plots for the number of goals scored by Brazil and Germany in regression \eqref{equ:independent-regression1}.}
\label{fig:regression-plot-attack}
\end{figure}

As two examples of interest, we sketch in Figure \ref{fig:regression-plot-attack} the results of the regression in  (\ref{equ:independent-regression1}) for Germany and Brazil. The  dots show the observed data (i.e, number of scored goals on the $y$-axis in dependence of the opponent's strength on the $x$-axis) and the  line is the estimated mean depending on the opponent's Elo strength.

\begin{figure}
\begin{center}
\includegraphics[width=6cm]{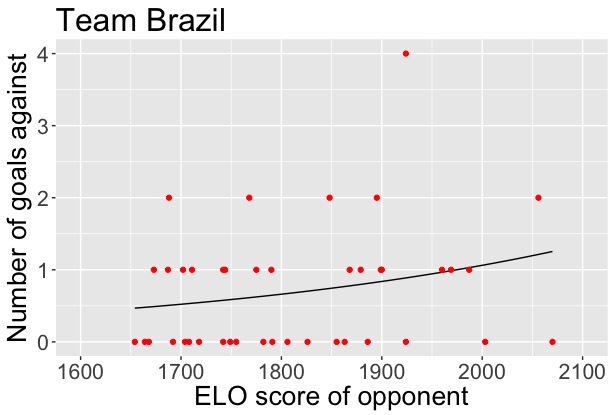}
\hfill
\includegraphics[width=6cm]{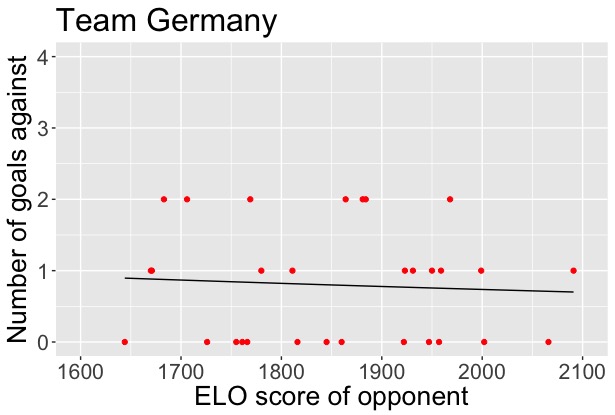} 
\end{center}
\caption{Plots for the number of goals against for  Brazil and Germany in regression \eqref{equ:independent-regression2}.}
\label{fig:regression-plot-defense}
\end{figure}

Analogously,   Figure \ref{fig:regression-plot-defense} sketches  the regression in  (\ref{equ:independent-regression2}) for Germany and Brazil. The  dots show the observed data (i.e., the number of goals against) and the  line is the estimated mean for the number of goals against.

\subsubsection{Goodness of fit test}\label{subsubsection:gof}
We check goodness of fit of  the Poisson regressions in (\ref{equ:independent-regression1}) and (\ref{equ:independent-regression2})  for all participating teams. For each team $\mathbf{T}$ we calculate the following  $\chi^{2}$-statistic from the list of matches:
$$
\chi_\mathbf{T} = \sum_{i=1}^{n_\mathbf{T}} \frac{(x_i-\hat\mu_i)^2}{\hat\mu_i},
$$
where $n_\mathbf{T}$ is the number of matches of team $\mathbf{T}$, $x_i$ is the number of scored goals of team $\mathbf{T}$ in match $i$ and $\hat\mu_i$ is the estimated Poisson regression mean. 
\par
\begin{figure}[ht]
\begin{center}
\includegraphics[width=6.0cm]{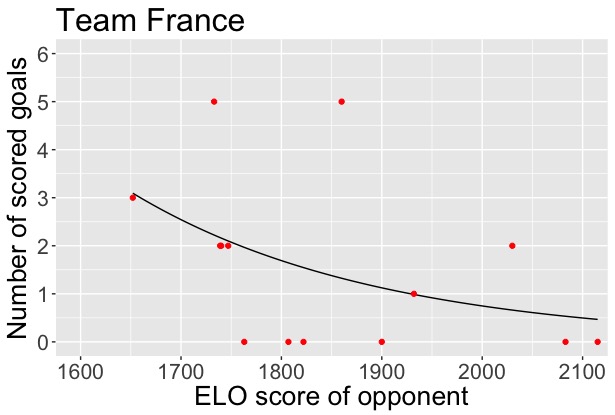} \quad\quad\quad
\includegraphics[width=6.0cm]{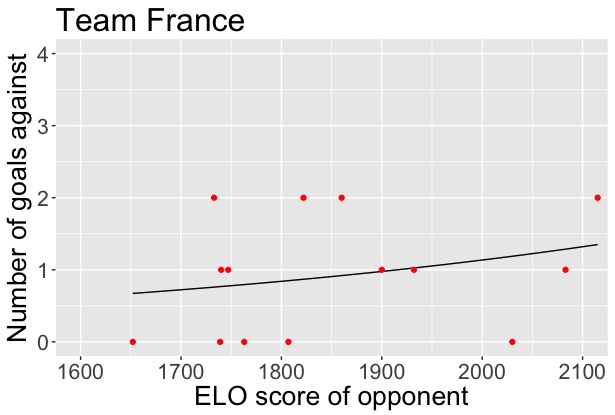}
\end{center}
\caption{Regression plots of France for the number of goals scored and goals against in regressions \eqref{equ:independent-regression1} and \eqref{equ:independent-regression2}. }
\label{fig:regression-france}
\end{figure}

We observe that most teams have a good fit, except some teams of less impact and France. We have found out that the bad fit of France is also a consequence of the bad and chaotic performance during the World Cup 2010. Therefore, we considered only French matches after 01.01.2012 but have also taken the matches of the EURO 2016 (held in France) into account  when fitting the parameters of France.
As an consequence the regression plots in Figure \ref{fig:regression-france} promise an acceptable fit.
The $p$-values for the top $5$ teams are given in Table \ref{table:godness-of-fit}. We remark that without the specific adaption for France we would get a $p$-value of $0.0011$ for France.
\begin{table}[H]
\centering
\begin{tabular}{l|c|c|c|c|c}
  \hline
 Team & Brazil & Germany & Spain & Argentina & France 
 \\ 
  \hline
  p-value & 0.56 &0.39 &  0.40 & 0.14  &0.03 
    \\
   \hline
\end{tabular}
\caption{Goodness of fit test for the independent Poisson regression model defined in Section \ref{subsec:independent-PR} for the top five teams. }
\label{table:godness-of-fit}
\end{table}

\subsubsection{Deviance analysis}
First, we calculate the null and residual deviances for each team for the regression in (\ref{equ:independent-regression1}). Table \ref{table:deviance-IndPR1} shows the deviance values and the $p$-values for the residual deviance for the top five teams in the current Elo ranking.  Although several of the $p$-values are low, they are still acceptable.
\begin{table}[ht]
\centering
\begin{tabular}{l|c|c|c}
  \hline
 Team & Null deviance & Residual deviance & p-value 
 \\ 
  \hline
 Brazil & 65.03 &         50.04 &                   0.21 \\ 
 Germany & 41.83 &  34.99 &   0.20\\  
   Spain & 54.83 &         38.89 &                     0.26\\ 
    Argentina & 59.65    &     47.19 &                    0.10\\ 
   France & 30.08 &          25.66 &                    0.019 \\ 
   \hline
\end{tabular}
\caption{Deviance analysis for the top five teams in  regressions   (\ref{equ:independent-regression1})}
\label{table:deviance-IndPR1}
\end{table}
The deviances and the $p$-values for the regression in  (\ref{equ:independent-regression2}) are given in Table \ref{table:deviance-IndPR2}. 
\begin{table}[h]
\centering
\begin{tabular}{l|c|c|c}
  \hline
  Team & Null deviance & Residual deviance & p-value
  \\ 
  \hline
 Brazil & 48.06 &         45.98 &                   0.35\\ 
 Germany & 31.69 &        31.48 &                    0.34\\ 
 Spain & 49.65 & 46.93 & 0.07 \\ 
   Argentina & 52.49 & 51.68 & 0.04 \\ 
   France & 14.81 &      13.50 &               0.41\\ 
     \hline
\end{tabular}
\caption{Deviance analysis for the top five teams in  regressions  (\ref{equ:independent-regression2})}
\label{table:deviance-IndPR2}
\end{table}


\subsection{Bivariate Poisson regression model}
The possible weakness of the previous model is that the number of goals $G_A$ and $G_B$ are realized independently and the fits in Tables \ref{table:deviance-IndPR1} and \ref{table:deviance-IndPR2} are not overwhelming. In this section we make a bivariate Poisson regression approach. First, recall the definition of a \textit{bivariate Poisson} distribution: let $X_1,X_2,X_0$ be \textit{independent} Poisson distributed random variables with rates $\lambda_1,\lambda_2,\lambda_0$. Define $Y_1=X_1+X_0$ and $Y_2=X_2+X_0$. Then $(Y_1,Y_2)$ is \textit{bivariate} Poisson distributed with parameters $(\lambda_1,\lambda_2,\lambda_0)$. In particular, $Y_i$ is Poisson distributed with rate $\lambda_i+\lambda_0$ and $\mathrm{Cov}(Y_1,Y_2)=\lambda_0$.
\par
The plan is to model $(G_A,G_B)$ as a bivariate Poisson distributed random vector for every couple $A,B$ separately (in order to keep each participants individual strengths). The main idea is two perform one regression over all matches of team $A$ and to estimate the average number of goals of team $A$ and its opponent in terms of his Elo strength $\elo{B}$. Then we  perform another regression over all matches of team $B$ and  estimate the expected number of goals of $B$ and the average goals against of $B$ when playing against a team of Elo rank $\elo{A}$. Hereby, we use the same notation as in Section \ref{subsec:independent-PR} with $\mu_{\mathbf{T}}$ being the Poisson rate for the average of scored goals of any team ${\mathbf{T}}$, while $\nu_{\mathbf{T}}$ is the average number of goals against of team ${\mathbf{T}}$.
The model  uses the following regression approach.
\begin{enumerate}
\item
For {each} World Cup participating team $\mathbf{T}$, we estimate the parameters $$(\lambda_1,\lambda_2, \lambda_0)=(\mu_{\mathbf{T}},\nu_{\mathbf{T}},\tau_{\mathbf{T}})$$  from the viewpoint of team $\mathbf{T}$, where we only take matches of team $\mathbf{T}$ on neutral playground into account. The parameters shall depend on the Elo strength $\elo{\O}$ of an opponent team $\O$. To this end, we use the following Poisson regression model:
\begin{eqnarray}
\log \mu_{\mathbf{T}}\bigl(\elo{\O}\bigr) &=& \alpha_{1,0} +\alpha_{1,1}\elo{\O},\nonumber\\
\log \nu_{\mathbf{T}}\bigl(\elo{\O}\bigr) &=& \alpha_{2,0} +\alpha_{2,1}\elo{\O},\label{equ:bivariate-regression1}\\
\log \tau_{\mathbf{T}}\bigl(\elo{\O}\bigr) &=& \alpha_{3,0} \nonumber
\end{eqnarray}
That is, the estimated expected number of scored goals of team $\mathbf{T}$ against a team of Elo strength $\elo{\O}$ is given by $\mu_{\mathbf{T}}\bigl(\elo{\O}\bigr)+\tau_{\mathbf{T}}$, while the estimated expected number of scored goals of a team with Elo score $\elo{\O}$ against $\mathbf{T}$ is given by $\nu_{\mathbf{T}}\bigl(\elo{\O}\bigr)+\tau_{\mathbf{T}}$. 
\par
\item In order to estimate the Poisson rates $(\lambda_1,\lambda_2,\lambda_0)$ for the match result $(G_A,G_B)$ we can use the regression coefficients both of $A$ and $B$ in the following way: 
 $\lambda_1$ may be estimated either by considering all matches of team $A$ and calculating $\mu_A\bigl(\elo{B}\bigr)$ or by considering all matches of team $B$ and calculating $\nu_B\bigl(\elo{A}\bigr)$, which corresponds to the goals against of team $B$ (that is, the number of scored goals of team $A$ against $B$). Therefore, we estimate $\lambda_1$ as the mean of $\mu_A\bigl(\elo{B}\bigr)$  and $\nu_B\bigl(\elo{A}\bigr)$. Analogously, we estimate $\lambda_2$ as the mean of $\mu_B\bigl(\elo{A}\bigr)$  and $\nu_A\bigl(\elo{B}\bigr)$ and $\lambda_0$ also as the mean of the covariances $\tau_A$ and $\tau_B$. That is,
\begin{eqnarray*}
\lambda_1 &=& \frac{\mu_A\bigl(\elo{B}\bigr)+ \nu_B\bigl(\elo{A}\bigr)}{2},\\
\lambda_2 &=& \frac{\mu_B\bigl(\elo{A}\bigr)+ \nu_A\bigl(\elo{B}\bigr)}{2},\\
\lambda_0 &=& \frac{ \tau_A\bigl(\elo{B}\bigr) +\tau_B\bigl(\elo{A}\bigr) }{2},\\
\end{eqnarray*}
\item Finally, we assume that $(G_A,G_B)$ is bivariate Poisson distributed with parameters $(\lambda_1,\lambda_2,\lambda_0)$.
\end{enumerate}

\textbf{Remark:} In (\ref{equ:bivariate-regression1}) we estimate $\tau_{\mathbf{T}}$ to be a constant for each team. Of course, one can also $\tau_{\mathbf{T}}$ let depend on the opponent's Elo score $\elo{\O}$, that is,
$$
\log \tau_{\mathbf{T}}\bigl(\elo{\O}\bigr) = \alpha_{3,0}+\alpha_{3,1}\elo{\O}.
$$
Calculations, however show that the AIC increases by adding the covariate $\elo{\O}$ for most of the teams. The same observation is made if we add $\elo{\mathbf{T}}$ as an additional covariate.


\subsection{Bivariate Poisson regression with diagonal inflation}
We consider the previous model with additional diagonal inflation. 
Such models are quite useful when one expects diagonal combinations with higher probabilities than the ones fitted under a bivariate Poisson model. In particular, it has been observed earlier, e.g. see \cite{KaNt:03}, \cite{KaNt:05}, that the number of draws is in some situation larger than those predicted by a simple bivariate Poisson model. 
We inflate the diagonal with probability $p$. The inflation is given by  the vector $(\theta_0,\theta_1,\theta_2)$ that describes the probability of the match results $0$:$0$, $1$:$1$ and $2$:$2$. We compare the AIC of the diagonal inflated model with the non-inflated model, see  Table \ref{table:inflated} for the five top teams.
The values of the inflation probability are close to zero. Despite the fact that the AIC decreases for almost all teams we do not believe that the inflated model improves the forecast. This observation is also supported by the results in Tables \ref{tab:score14} and \ref{tab:score10}.
\begin{table}[h]
\centering
\begin{tabular}{lcccccc}
  \hline
  Team & $p$ & $\theta_{0}$ & $\theta_{1}$ & $\theta_{2}$ & AIC & AIC  \\ 
 	&	&	&	&	& inflated	& not inflated\\
  \hline
 Brazil & 0.01 & 0.00 & 0.00 & 1.00 & \textbf{251.56} & 257.40 \\ 
  Germany & 0.01& 0.00 & 0.00 & 1.00 & \textbf{186.19} &        192.18\\  
   Spain & 0.00 & 0.00 & 0.00 & 1.00 & \textbf{215.06} & 221.06 \\ 
  Argentina & 0.02 & 0.00 & 0.00 & 1.00 & \textbf{230.97} & 236.56 \\ 
   France & 0.03 & 1.00 & 0.00 &0.00 &  \textbf{93.53} &         99.46\\  
   \hline
\end{tabular}
\caption{Diagonal inflated bivariate Poisson regression for the top five teams. }
\label{table:inflated}
\end{table}

\subsection{Nested Poisson regression model}

 We now present another \textit{dependent} Poisson regression approach. The Poisson rates $\lambda_{A|B}$ and $\lambda_{B|A}$ are now determined as follows: 
\begin{enumerate}
\item We always assume that $A$ has higher Elo score than $B$. This assumption can be justified, since usually the better team dominates the weaker team's tactics. Moreover the number of goals the stronger team scores has an impact on the number of goals of the weaker team. For example,  if team $A$ scores  $5$ goals it is more likely that $B$ scores also $1$ or $2$ goals, because the defense of team $A$ lacks in concentration  due to the expected victory. If the stronger team $A$ scores only $1$ goal, it is more likely that $B$ scores no or just one goal, since team $A$ focusses more on the defence  and secures the victory.

\item The Poisson rate for $G_A$ is determined as in Section \ref{subsec:independent-PR} before by
$$
\lambda_{A|B} = \frac{\mu_A\bigl(\elo{B}\bigr)+\nu_B\bigl(\elo{A}\bigr)}{2},
$$
which is obtained via Poisson regression.
\item The number of goals $G_B$ scored by $B$ is assumed to depend on the Elo score $E_A=\elo{A}$ and additionally on the outcome of $G_A$. More precisely, $G_B$ is modeled as a Poisson distribution with parameter $\lambda_B(E_A,G_A)$ satisfying
\begin{equation}\label{equ:nested-progression-2}
\log \lambda_B(E_A,G_A) = \gamma_0 + \gamma_1 \cdot E_A+\gamma_2 \cdot G_A.
\end{equation}
Once again, the parameters $\gamma_0,\gamma_1,\gamma_2$ are obtained by Poisson regression. Hence,
$$
\lambda_{B|A} = \lambda_B(E_A,G_A).
$$
\item The result of the match $A$ versus $B$ is simulated by realizing $G_A$ first and then  realizing $G_B$ in dependence of the realization of $G_A$. 
\end{enumerate}
This approach may also be justified through the definition of  conditional probabilities:
$$
\mathbb{P}[G_A=i,G_B=j] = \mathbb{P}[G_A=i]\cdot \mathbb{P}[G_B=j \mid G_A=i] \quad \forall i,j\in\mathbb{N}_0.
$$
We are not aware of other validation methods for this model than the validation on historical data. Tables \ref{tab:score14} and \ref{tab:score10} indicate that this model may indeed have the best fit.




\section{Score functions}\label{sec:scores}

In the following we want to compare the predictions with actual results of the two previous FIFA World Cups. For this purpose, we introduce the following notation. For a team  $\mathbf{T}$ we define:
$$
\mathrm{result}(\mathbf{T}) = \begin{cases}
1, &\textrm{if } \mathbf{T} \textrm{ was FIFA World Cup  winner}, \\
2, &\textrm{if } \mathbf{T} \textrm{ lost the final},\\
3, &\textrm{if } \mathbf{T} \textrm{ dropped out in semifinal},\\
4, &\textrm{if } \mathbf{T} \textrm{ dropped out in quarterfinal},\\
5, &\textrm{if } \mathbf{T} \textrm{ dropped out in round of last 16},\\
6, &\textrm{if } \mathbf{T} \textrm{ dropped out in round robin}.
\end{cases}
$$
For example, in 2014 we have $\mathrm{result(Germany)}=1$, $\mathrm{result(Argentina)}=2$, $\mathrm{result(Brazil)}=3$ or $\mathrm{result(Italy)}=6.$ We consider the variable $\mathrm{result}$ as a ordinal variable, since for instance predicting Germany to drop out in round robin should be penalized more than predicting that Germany looses the final. We choose a linear scaling, i.e. the values $1,2,3,4,5,6$, since there is always one match between the different rounds.  Score functions for ordinal variables are, to the best of our knowledge, not well studied. We refer to \cite{CoFe:12} for a discussion on this topic. We propose two new score functions and compare them with the Brier score and the Rank-Probability-Score (RPS). For each model, the simulation leads to a probability distribution given by $p_j(\mathbf{T})=\mathbb{P}[ \mathrm{result}(\mathbf{T})=j]$, $j\in\{1,\dots,6\}$, for the result of each team $\mathbf{T}$. The following score functions measure and compare the forecasts with the real outcome.

\begin{enumerate}
\item \textbf{Maximum-Likelihood-Score:} The error of team $\mathbf{T}$ is  defined as
$$
\mathrm{error}_{1}(\mathbf{T}) = \bigl| \mathrm{result}(\mathbf{T}) - \underset{i=1,\dots,6}{\operatorname{argmax}}~ p_i(\mathbf{T})]\bigr|.
$$
The total error score is given by summing up the errors of all World Cup participating teams:
$$
E_1 = \sum_{\mathbf{T}} \mathrm{error}_{1}(\mathbf{T}). 
$$
\item \textbf{Weighted differences:}
The error of team $\mathbf{T}$ is  defined as
$$
\mathrm{error}_2(\mathbf{T}) = \sum_{j=1}^6  p_{j}(\mathbf{T}) \bigl| j-\mathrm{result}(\mathbf{T}) \bigr|.
$$
The total error score is  given by
$$
E_2  = \sum_{\mathbf{T}} \mathrm{error}_2(\mathbf{T}).
$$ 
\item \textbf{Brier Score:}
The error of team $\mathbf{T}$ is defined as
$$
\mathrm{error}(\mathbf{T})_{3} = \sum_{j=1}^6 \bigl( p_{j}(\mathbf{T}) -\mathds{1}_{[\mathrm{result}(\mathbf{T})=j]}  \bigr)^2.
$$
The total error score is  given by
$$
BS  = \sum_{\mathbf{T}} \mathrm{error}_3(\mathbf{T}).
$$ 
\item \textbf{Rank-Probability-Score (RPS):} 
The error of team $\mathbf{T}$ is  defined as
$$
\mathrm{error}_{4}(\mathbf{T}) = \frac{1}{5} \sum_{i=1}^5 \left( \sum_{j=1}^i p_{j}(\mathbf{T}) -\mathds{1}_{[\mathrm{result}(\mathbf{T})=j]}  \right)^2.
$$
The total error score is  given by
$$
RPS  = \sum_{\mathbf{T}} \mathrm{error}_{4}(\mathbf{T}). 
$$ 

\end{enumerate}

We note that the dependence of the different outcomes of the teams penalizes exceptional outcomes as underdogs wins and early drop outs of favorites. 

\section{Validation of Models on FIFA World Cup 2014 results}
\label{sec:val14}

In this section we test the different models from the previous section on the FIFA World Cup 2014 data. For this purpose, we take into account all matches between 01.01.2002 and the beginning of the tournament of all participants on \textit{neutral playground}. We remark that it was necessary to take historical match data up to $12$ years before the tournament due to lack of enough matches for reasonable fits (e.g, the regression for Belgium matches was not satisfying). We simulate the whole tournament according to the FIFA rules, that is, at the end of the group stage the final group table is evaluated according to  the FIFA rules (except Fair-Play criterion). Additional, after each game the Elo scores of the teams are updated. Furthermore, the score of a match which goes into extra time is simulated with the same Poisson rates as for a match of $90$ minutes but with rates divided by $3$ (extra time is $30$ minutes $= 90$ minutes $/3$). For each model $100.000$ simulations are performed. We have implemented the simulation in \texttt{R} version 3.3.1, where we used the \texttt{bivpois}-package of Karlis and Ntzoufras, which uses the EM algorithm for estimating the parameters.

The simulation results are given in the following form. For each team we estimate the probability that the team reaches a certain  round or wins the tournament. For instance, 
\begin{center}
\begin{tabular}{lcccccc}
  \hline
   Team & World Champion & Final &  Semi & Quarter & R16 & Prelim. Round \\ 
  \hline
   Brazil & 20.30 & 30.30 & 40.30 & 54.80 & 86.10 & 13.90 \\ 
\end{tabular}
\end{center}
means that Brazil wins the cup with a probability of $20.30\%$, reaches the final with probability $30.30\%$, the semifinals with probability $40.30\%$ etc. The last column gives the probability to drop out in the group phase.
For each team we mark in bold type the   round where the team actually  dropped out.  Results for independent Poisson regression and the nested regression model  may be found in Tables  \ref{table:ind14} and  \ref{tab:nested14}. The tables for the other models can be found in Appendix \ref{app:14}.


\begin{table}[H]
\centering
\begin{tabular}{llcccccc}
  \hline
 & Team & World Champion & Final &  Semi & Final & R16 & Prelim. Round \\ 
  \hline
1 & Spain & 21.80 & 32.50 & 42.20 & 58.40 & 88.40 & \textbf{11.60} \\ 
  2 & Brazil & 20.30 & 30.30 & \textbf{40.30} & 54.80 & 86.10 & 13.90 \\ 
  3 & Germany & \textbf{11.90} & 23.80 & 51.20 & 69.70 & 85.20 & 14.80 \\ 
  4 & Netherlands & 7.70 & 14.60 & \textbf{23.10} & 37.50 & 71.50 & 28.50 \\ 
  5 & Portugal & 5.50 & 12.90 & 30.10 & 48.10 & 67.90 & \textbf{32.10} \\ 
  6 & Argentina & 5.10 & \textbf{13.90} & 36.90 & 65.60 & 92.10 & 7.90 \\ 
  7 & England & 4.10 & 9.40 & 17.20 & 43.30 & 69.20 & \textbf{30.80} \\ 
  8 & Uruguay & 3.80 & 8.50 & 15.90 & 40.10 & \textbf{66.70} & 33.30 \\ 
  9 & Italy & 2.20 & 5.20 & 10.40 & 28.10 & 52.20 & \textbf{47.80} \\ 
  10 & Russia & 2.10 & 5.70 & 15.30 & 29.30 & 73.80 & \textbf{26.20} \\ 
  11 & Colombia & 1.90 & 4.70 & 9.80 & \textbf{27.80} & 59.20 & 40.80 \\ 
  12 & France & 1.90 & 4.70 & 12.10 & \textbf{28.50} & 57.80 & 42.20 \\ 
\hline
\end{tabular}
\caption{FIFA World Cup 2014 prediction via independent Poisson regression}
\label{table:ind14}
\end{table}

\begin{table}[H]
\centering
\begin{tabular}{llcccccc}
  \hline
 & Team & World Champion & Final &  Semi & Quarter & R16 & Prelim. Round \\ 
  \hline
1 & Brazil & 20.00 & 28.50 &  \textbf{36.80} & 51.40 & 81.30 & 18.70 \\ 
  2 & Spain & 16.70 & 26.20 & 35.10 & 51.50 & 85.10 &  \textbf{14.90} \\ 
  3 & Germany &  \textbf{13.50} & 25.10 & 51.10 & 67.60 & 84.50 & 15.50 \\ 
  4 & Netherlands & 8.80 & 15.90 &  \textbf{24.80} & 40.10 & 74.10 & 25.90 \\ 
  5 & Argentina & 8.10 &  \textbf{19.30} & 44.00 & 71.50 & 92.90 & 7.10 \\ 
  6 & Uruguay & 6.00 & 12.20 & 21.80 & 47.60 &  \textbf{73.20} & 26.80 \\ 
  7 & England & 4.20 & 9.50 & 18.00 & 43.70 & 69.60 &  \textbf{30.40} \\ 
  8 & Russia & 3.20 & 7.70 & 17.80 & 32.70 & 70.90 &  \textbf{29.10} \\ 
  9 & Portugal & 2.70 & 8.20 & 21.70 & 38.60 & 58.20 &  \textbf{41.80} \\ 
  10 & Colombia & 2.00 & 4.90 & 10.20 &  \textbf{27.40} & 61.10 & 38.90 \\ 
  \hline
\end{tabular}
\caption{FIFA World Cup 2014 prediction via nested Poisson regression}
\label{tab:nested14}
\end{table}

The Elo ratings as they were  on 11 june 2014  for the top $5$ nations (in this rating) are as follows:
\begin{center}
\begin{tabular}{c|c|c|c|c}
\hline
Brazil & Spain & Germany & Argentina & Netherlands \cr
\hline
2113 & 	2086    & 	2046  & 1989	& 1959 \cr
\hline
\end{tabular}
\end{center}
We clearly see that the forecasts correspond roughly to the Elo ranking, but the different modeling of team specific strengths changes the ordering of the teams slightly. Moreover, certain probabilities differ significantly in the different models. For example, Spain wins the cup with probability $21.80\%$ in the independent regression model but only with probability $16.70\%$ in the nested regression model. We also see that the early drop out of Italy was not so surprising after all.

In  Table \ref{tab:score14} we compare the different models by calculating the different scores from Section \ref{sec:scores}. The nested Poisson regression scores best for the $E_{1}$ score, Brier score and RPS. The $E_{2}$ score favors slightly the bivariate model.
\begin{table}[h]
\centering
\begin{tabular}{lccccccc}
  \hline
Models  & $E_1$  & $E_2$  & Brier & RPS\\
\hline
Independent Poisson regression & 26 & 34.65  & 22.10 & 5.48\\
Nested Poisson regression  &\textbf{25}& {34.32} & \textbf{21.89} & \textbf{5.42}\\
Bivariate Poisson regression  & 28 & \textbf{34.16}  & 22.33  & 5.52 \\
Diagonal Inflated Bivariate Poisson regression  & 26 &  34.68  & 22.13 & 5.48 \\
  \hline
\end{tabular}
\caption{Scores for the FIFA World Cup 2014 simulations}
\label{tab:score14}
\end{table}

\section{Validation of Models on FIFA World Cup 2010 results}
The simulations for the FIFA World Cup 2010 are done as in the previous section, but using data between 01.01.2000  and the beginning of the championship.

Results for the independent regression model and the nested regression model are given in Tables \ref{tab:indep10} and \ref{tab:nested10}. The tables for the other models can be found in Appendix  \ref{app:10}.

\begin{table}[H]
\centering
\begin{tabular}{rlcccccc}
  \hline
 & Team & World Champion & Final & Semi & Quarter & R16 & Prelim. Round \\ 
  \hline
1 & Brazil & 25.50 & 34.10 & 45.10 & \textbf{61.60} & 78.60 & 21.40 \\ 
  2 & Netherlands & 15.00 & \textbf{24.40} & 36.40 & 67.30 & 87.70 & 12.30 \\ 
  3 & Spain & \textbf{11.20} & 23.00 & 32.90 & 49.80 & 90.90 & 9.10 \\ 
  4 & England & 7.80 & 15.50 & 34.30 & 57.80 & \textbf{85.10} & 14.90 \\ 
  5 & Portugal & 5.80 & 13.40 & 22.70 & 38.60 & \textbf{69.10} & 30.90 \\ 
  6 & Italy & 4.80 & 11.20 & 19.90 & 48.40 & 86.90 & \textbf{13.10} \\ 
  7 & France & 4.80 & 9.00 & 18.00 & 31.50 & 58.40 & \textbf{41.60} \\ 
  8 & Argentina & 3.50 & 10.50 & 28.70 & \textbf{51.80} & 86.60 & 13.40 \\ 
  9 & Uruguay & 3.30 & 6.80 & \textbf{15.20} & 28.80 & 55.80 & 44.20 \\ 
  10 & South Korea & 2.70 & 5.40 & 11.40 & 22.40 & \textbf{45.20} & 54.80 \\ 
  11 & Germany & 2.30 & 9.80 & \textbf{33.60} & 64.80 & 93.00 & 7.00 \\ 
   \hline
\end{tabular}
\caption{FIFA World Cup 2010 prediction via independent Poisson regression}
\label{tab:indep10}
\end{table}

\begin{table}[H]
\centering
\begin{tabular}{rlcccccc}
  \hline
 & Team & World Champion & Final &  Semi & Quarter & R16 & Prelim. Round \\ 
  \hline
1 & Brazil & 19.70 & 27.60 & 38.40 & \textbf{54.90} & 69.30 & 30.70 \\ 
  2 & Spain & \textbf{14.70} & 26.80 & 36.80 & 51.80 & 94.40 & 5.60 \\ 
  3 & Netherlands & 14.60 & \textbf{23.90} & 37.30 & 65.10 & 87.60 & 12.40 \\ 
  4 & England & 7.00 & 14.60 & 32.80 & 57.80 & \textbf{86.20} & 13.80 \\ 
  5 & Uruguay & 6.40 & 11.70 & \textbf{21.50} & 36.60 & 62.90 & 37.10 \\ 
  6 & France & 4.70 & 8.70 & 16.20 & 28.50 & 52.30 & \textbf{47.70} \\ 
  7 & Portugal & 4.30 & 10.10 & 18.40 & 32.20 & \textbf{60.50} & 39.50 \\ 
  8 & Italy & 4.00 & 9.80 & 19.70 & 50.90 & 89.80 & \textbf{10.20} \\ 
  9 & South Korea & 3.90 & 7.10 & 13.20 & 25.50 & \textbf{52.40} & 47.60 \\ 
  10 & Argentina & 3.70 & 9.80 & 26.50 & \textbf{48.00} & 84.20 & 15.80 \\ 
\hline
\end{tabular}
\caption{FIFA World Cup 2010 prediction via nested Poisson regression}
\label{tab:nested10}
\end{table}

The Elo ratings as they were  on 11 june 2010  for the top $5$ nations (in this rating) are as follows:
\begin{center}
\begin{tabular}{c|c|c|c|c}
\hline
Brazil & Spain & Netherlands & England & Germany \cr
\hline
 2087 & 	2085    & 	2016  & 1975	&  1929 \cr
\hline
\end{tabular}
\end{center}

 The early drop out of Brazil may be unlikely at first sight, but at second sight seems not so surprising since 
they lost against the Netherlands in the quarter final.  The forecast of Germany becoming world champion is remarkably low in all models. This reflects the influence of the course of the tournament since Germany played against England in R16, Argentina in the quarters and Spain in the semi finals. Interesting to note that the early drop out of France had already a high probability. Knowing this ``le fiasco de Knysna'' may have been avoided.

The error scores for the World Cup 2010 tournament are given in Table \ref{tab:score10}. Again the nested Poisson regression scores best, now even for all four score functions.
 \begin{table}[h]
\centering
\begin{tabular}{lccccccc}
  \hline
Models  & $E_1$  & $E_2$  & Brier &RPS \\
\hline
Independent Poisson regression  & 25  &   30.97     &17.97 & 5.05\\
Nested Poisson regression  & \textbf{24}  & \textbf{30.50} & \textbf{17.51} &  \textbf{4.93} \\
Bivariate Poisson regression  & 25 &   33.04     & 17.97 & {4.99} \\
Diagonal Inflated Bivariate Poisson regression  & 25  & 31.49  & 17.96 &  5.08 \\
  \hline
\end{tabular}
\caption{Scores for FIFA World Cup 2010}
\label{tab:score10}
\end{table}


\textbf{Remarks:}
In order to avoid a degenerate behaviour and a bad fit of our models we had to do the following adaptions:
\begin{itemize}
\item For the nested Poisson regression model for Slovenia only matches against other participants were used. 
\item For the  bivariate Poisson regression model for Germany, we have taken into account also the matches of Germany during the World Cup 2006 (held in Germany).
\item The matches of Serbia include also the matches of former ``Yugoslavia'' and ``Serbia and Montenegro'' before the year $2006$, in which Serbia started its own national team. 
\end{itemize}

\section{World Cup 2018 Simulations}

We simulate the whole tournament $100.000$ times for each model as described in Section \ref{sec:val14}. The models are fitted on matches of all participants on neutral ground since 01.01.2010. For France we consider only matches after 01.01.2012 and include also the matches of the EURO 2016; compare with Section \ref{subsubsection:gof}. 
 The results are presented in Tables \ref{tab:indep18},  \ref{tab:nested18}, \ref{tab:biv18} and \ref{tab:inflated18}. 
 The regression models coincide in the order of the first four favorites for the cup. In particular, they favor Germany and not Brazil. However, single probabilities may be quite different. For instance, Germany is estimated to win the cup with $26.00\%$ in the independent Poisson regression model and with  $30.50\%$ in the nested Poisson regression model.
The circumstance that all models do favor Germany and not Brazil may depend on team specific effects and the following fact. If both Germany and Brazil win their group they will meet only in the final. Now, if Germany reaches the final it more likely won against stronger teams and therefore is likely to have higher Elo ranking in the final than Brazil. This underlines the importance of the dynamic Elo updating in the simulations.


\begin{table}[ht]
\centering
\begin{tabular}{rlccccccc}
  \hline
 & Team & World Champion & Final &Semi & Quarter & R16 & Prelim. Round \\ 
  \hline
 1 & Germany & 26.00 & 36.50 & 52.10 & 68.80 & 92.50 & 7.40 \\ 
  2 & Brazil & 13.20 & 26.00 & 41.00 & 57.70 & 88.30 & 11.70 \\ 
  3 & Spain & 11.20 & 21.30 & 41.50 & 68.60 & 84.90 & 15.30 \\ 
  4 & Argentina & 9.20 & 16.70 & 31.90 & 53.60 & 84.50 & 15.50 \\ 
  5 & Colombia & 7.00 & 13.20 & 24.10 & 49.70 & 75.10 & 24.90 \\ 
  6 & Portugal & 5.90 & 13.30 & 28.80 & 53.80 & 73.90 & 26.20 \\ 
  7 & France & 5.30 & 12.40 & 26.00 & 46.60 & 79.70 & 20.30 \\ 
  8 & Peru & 4.30 & 9.20 & 19.00 & 35.80 & 67.70 & 32.30 \\ 
  9 & Belgium & 3.60 & 9.40 & 19.50 & 48.20 & 85.40 & 14.80 \\ 
  10 & Poland & 2.80 & 6.30 & 13.70 & 33.20 & 59.80 & 40.20 \\ 
      \hline
\end{tabular}
\caption{World Cup 2018 prediction via independent Poisson regression}
\label{tab:indep18}
\end{table}


\begin{table}[ht]
\centering
\begin{tabular}{rlccccccc}
  \hline
 & Team & World Champion & Final &  Semi & Quarter & R16 & Prelim. Round \\ 
  \hline
 1 & Germany & 30.50 & 41.80 & 57.70 & 74.30 & 92.10 & 7.90 \\ 
  2 & Brazil & 18.30 & 33.60 & 46.20 & 61.90 & 93.20 & 6.70 \\ 
  3 & Spain & 13.90 & 24.80 & 47.50 & 70.80 & 90.00 & 10.10 \\ 
  4 & Argentina & 8.30 & 16.20 & 32.30 & 57.10 & 86.20 & 13.80 \\ 
  5 & Colombia & 4.30 & 9.40 & 19.30 & 45.00 & 73.40 & 26.60 \\ 
  6 & Portugal & 3.90 & 11.00 & 27.40 & 52.00 & 75.90 & 24.10 \\ 
  7 & France & 3.40 & 10.10 & 25.10 & 46.70 & 78.90 & 21.20 \\ 
  8 & Belgium & 3.00 & 8.70 & 18.70 & 50.40 & 83.10 & 16.80 \\ 
  9 & Russia & 2.80 & 5.30 & 10.40 & 21.50 & 49.00 & 51.00 \\ 
  10 & England & 2.60 & 6.70 & 15.30 & 43.60 & 75.80 & 24.40 \\ 
  \hline
\end{tabular}
\caption{World Cup 2018 prediction via nested Poisson regression}
\label{tab:nested18}
\end{table}


\begin{table}[ht]
\centering
\begin{tabular}{rlccccccc}
  \hline
 & Team & World Champion & Final & Semi & Quarter & R16 & Prelim. Round \\ 
  \hline
  1 & Germany & 26.90 & 37.30 & 52.30 & 70.70 & 93.20 & 6.80 \\ 
  2 & Brazil & 13.00 & 26.00 & 40.10 & 59.80 & 89.70 & 10.30 \\ 
  3 & Spain & 11.20 & 21.20 & 41.70 & 69.20 & 86.20 & 13.90 \\ 
  4 & Argentina & 9.60 & 17.30 & 33.60 & 56.30 & 88.20 & 11.90 \\ 
  5 & Colombia & 8.40 & 15.30 & 27.80 & 59.10 & 82.20 & 17.80 \\ 
  6 & France & 5.20 & 12.60 & 27.90 & 49.20 & 81.80 & 18.30 \\ 
  7 & Portugal & 5.20 & 13.00 & 28.90 & 53.70 & 76.00 & 24.10 \\ 
  8 & Belgium & 3.90 & 10.60 & 22.40 & 55.90 & 89.70 & 10.40 \\ 
  9 & Peru & 3.90 & 8.80 & 19.00 & 36.00 & 68.50 & 31.40 \\ 
  10 & England & 2.10 & 4.90 & 11.70 & 31.40 & 76.70 & 23.20 \\ 
      \hline
\end{tabular}
\caption{World Cup 2018 prediction via bivariate Poisson regression}
\label{tab:biv18}
\end{table}


\begin{table}[ht]
\centering
\begin{tabular}{rlccccccc}
  \hline
 & Team & World Champion & Final & Semi & Quarter & R16 & Prelim. Round \\ 
  \hline
 1 & Germany & 25.80 & 36.10  & 51.00 & 68.70 & 91.40 & 8.60 \\ 
  2 & Brazil & 12.30 & 24.30  & 38.50 & 57.00 & 89.60 & 10.30 \\ 
  3 & Spain & 11.20 & 20.90  & 40.40 & 66.60 & 85.70 & 14.30 \\ 
  4 & Argentina & 9.60 & 17.20  & 33.00 & 55.00 & 86.70 & 13.10 \\ 
  5 & Colombia & 7.60 & 14.10  & 26.30 & 55.50 & 81.00 & 18.90 \\ 
  6 & Portugal & 5.20 & 12.40  & 27.60 & 51.70 & 75.60 & 24.30 \\ 
  7 & France & 5.00 & 11.60  & 24.80 & 44.30 & 75.40 & 24.60 \\ 
  8 & Belgium & 3.90 & 10.50  & 21.90 & 53.90 & 87.10 & 12.90 \\ 
  9 & Peru & 3.70 & 8.50 & 18.20 & 35.20 & 67.50 & 32.50 \\ 
  10 & England & 2.20 & 4.60  & 10.00 & 26.40 & 59.30 & 40.70 \\ 
   \hline
\end{tabular}
\caption{World Cup 2018 prediction via diagonal inflated Poisson regression}
\label{tab:inflated18}
\end{table}


\section{Sankey}
\begin{figure}[ht]
\begin{center}
\includegraphics[angle=90, width=13.0cm]{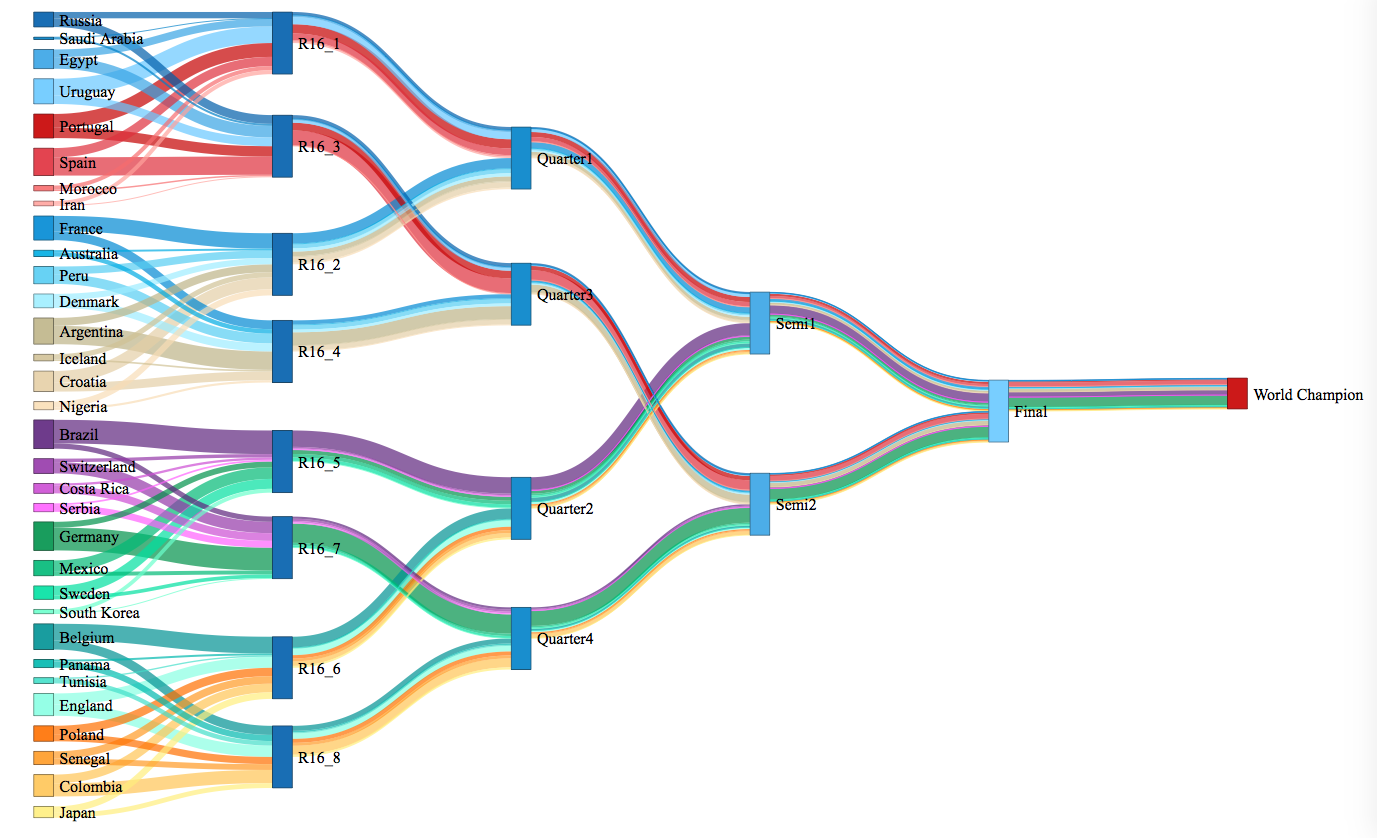} 
\end{center}
\caption{Sankey presentation of the forecast of the FIFA Worldcup 2018 based on 100.000 simulations of the nested regression model. }
\label{fig:sankey}
\end{figure}

We present the simulation results of the nested Poisson regression model in a Sankey diagram, see Figure \ref{fig:sankey}. The width of the edges correspond to the probabilities  of reaching stages in the the tournament.

\section{Discussion}
\label{sec:discussion}

In this section we want to give some quick discussion about the used Poisson models and related models. Of course, the Poisson models we used are not the only natural candidates for modeling football matches. Multiplicative mixtures may lead to overdispersion. Thus, it is desirable to use models having a variance function which is flexible enough to deal with overdispersion and underdispersion. One natural model for this is the \textit{generalised Poisson model}, which was suggested by  \cite{Co:89}. We omit the details
but remark that this distribution has an additional parameter $\varphi$ which allows to model the variance as $\lambda/\varphi^2$; for more details on generalised Poisson regression we refer to  \cite{St:04} and  \cite{Er:06}. Estimations of $\varphi$ by generalised Poisson regression lead to the observation that $\varphi$ is close to $1$ for the most important teams. Therefore, no additional gain is given by the use of the generalised Poisson model.  
\par
Another related candidate for the simulation of football matches is given by the \textit{negative binomial distribution}, where also another parameter comes into play to allow a better fit. However, the same observations as in the case of the generalised Poisson model can be made, that is, the estimates of the  additional parameter lead to a model which is almost just a simple Poisson model. We refer to  \cite{JoZh:09} for a detailed comparison of generalized Poisson distribution and negative Binomial distribution.
\par
A potential problem may rely in the fact that there are not sufficiently many matches for each team during the last eight years. This relies, in particular, on the fact that we considered only matches on \textit{neutral} playground. Of course, it is possible to include also matches from the qualifier rounds for the international tournaments. We followed this approach and, in order to weight home advantages/away disadvantages, we introduced another categorical covariate $L$ for the ``home advantage'' which lead to the following regression model  extending (\ref{equ:independent-regression1}) and (\ref{equ:independent-regression2}):
\begin{eqnarray*}
\log \mu_A(\elo{\O}) &= & \alpha_0 + \alpha_1 \cdot \elo{\O} + \alpha_2 \cdot L,\\
\log \nu_B(\elo{\O}) & = &\beta_0 + \beta_1 \cdot \elo{\O}+\beta_2\cdot L,
\end{eqnarray*}
where $L=1$ if $A$ plays at home, $L=-1$ if $B$ plays at home, and $L=0$ if the match is on neutral playground.
\par
Using this regression approach leads, however, to effects that hide the team's real strength in tournaments. In numbers, almost every team has then a probability between $2\%$ and $6\%$ of winning the World Cup, which obviously makes no sense. This in turn leads to the conclusion that matches during championships behave different than typical matches in the qualifier round.

We did not study the robustness of our models rigorously. However, we observed that the regressions models tend to be rather sensitive to the choices of matches. In particular, we had to adapt the time range of historical match data before each of the different World Cup simulations for 2010, 2014 and 2018. Although $8$ years seem to be a quite reasonable time range, it did not lead to  satisfying regression parameters for the World Cup 2010 and 2014 simulations. This explains why we had to take different time ranges for each World Cup under consideration.
We refer to    \cite{KaNt:11} for a detailed discussion on robustness. 
We note that the model is sensitive to changes of the Elo points during the tournament. Simulations where the Elo points are not updated during the tournament lead to quite different probabilities (up to $5$ percent points) and clearly favor the stronger teams. I particular, this shows that the dynamic Elo updating models the effect of alleged underdogs having a good run.

There have been attempts to improve the Elo rating. For instance,  
  \cite{CoFe:13} propose a dynamic rating that takes into account the relative ability between adversaries. In \cite{CoFe:13} it is shown that this rating outperforms in certain cases models based solely on Elo scores.  We also refer to \cite{CoFe:12b}, \cite{CoFe:13b} and references therein  for more details on Bayesian models for forecasting football  matches outcomes and to  \cite{HiWr:03} for Markov models of team specific characteristics.
The models above are in our opinion more appropriate for short term forecasts. As in our case we are interested in long term forecasts, random effects are of considerable influence, and we suspect that more sophisticated models do not a priori improve the quality of the forecast.
It goes without saying that a more intensive study on which data is relevant for (longtime) football forecasts is needed, e.g.  see   \cite{CoFe:17}.

Measuring the accuracy of any forecasting model is a critical part of its validation. In the absence of an agreed and appropriate type of scoring rule it is rather difficult to reach a consensus about  whether a model is sufficiently ``good'' or which of several different models is ``best''.  Our results show that the four scoring rules under considerations agree on the ``best'' model.  With the relentless increase in football forecasting sports events and tournaments  it will become more and more important to use  effective scoring rules for ordinal variables. Although we are not suggesting (neither are convinced) that our proposed scoring rules $E_{1}$ and $E_{2}$ and the  RPS are the only valid candidates for such a scoring rule, we have shown that they mostly,  at least in our setting, give the same result on which model is ``best''.

\section{Conclusion}
Several team-specific Poisson regression models for the number of goals in football matches  facing each other in international tournament matches are studied and compared. They all include the Elo points of the teams as covariates and use all FIFA matches of the teams since 2010 as underlying data.The fitted models were used for Monte-Carlo simulations  of the FIFA Worldcup 2018.  According to these simulations, Germany (followed by Brazil) turns out to be the top favorite for winning the title. Besides, for every team probabilities of reaching the different stages of the cup are calculated.

A major part of the statistical novelty of the presented work lies in the   introduction of two new score functions for ordinal variables as well as the construction of the nested regression model. This model  outperforms previous studied models, that use  (inflated) bivariate Poisson regression, when tested on the previous FIFA World Cups 2010 and 2014.  

We propose a weighted visualization of the course of the tournament using a large Sankey diagram. It enables experts and fans to obtain at a glance a quantified estimation of all kind of possible events.

\section{Appendix}
\subsection{FIFA World Cup 2014 simulations}\label{app:14}
We present the tables of the forecast for the FIFA World Cup 2014 simulations using the bivariate Poisson regression model and the diagonal inflated bivariate Poisson regression model. Each model was simulated $100.000$ times.

\begin{table}[h!]
\centering
\begin{tabular}{llcccccc}
  \hline
 & Team & World Champion & Final &  Semi & Quarter & R16 & Prelim. Round \\ 
  \hline
1 & Spain & 24.70 & 35.80 & 46.00 & 60.30 & 89.90 & \textbf{10.10} \\ 
  2 & Brazil & 21.00 & 31.20 & \textbf{41.50} & 54.90 & 86.90 & 13.10 \\ 
  3 & Germany & \textbf{11.80} & 24.20 & 52.40 & 70.40 & 85.70 & 14.30 \\ 
  4 & Netherlands & 7.80 & 15.00 & \textbf{23.70} & 37.70 & 72.50 & 27.50 \\ 
  5 & Portugal & 5.30 & 12.90 & 30.80 & 48.50 & 68.00 & \textbf{32.00} \\ 
  6 & Argentina & 4.80 & \textbf{13.90} & 38.00 & 68.10 & 93.80 & 6.20 \\ 
  7 & Uruguay & 3.60 & 8.30 & 15.80 & 41.20 & \textbf{67.50} & 32.50 \\ 
  8 & England & 3.00 & 8.00 & 15.00 & 44.70 & 70.60 & \textbf{29.40} \\ 
  9 & Russia & 2.20 & 6.30 & 16.90 & 32.60 & 79.60 & \textbf{20.40} \\ 
  10 & Italy & 2.10 & 4.90 & 9.90 & 27.20 & 50.80 & \textbf{49.20} \\ 
  11 & Colombia & 1.80 & 4.50 & 9.50 & \textbf{27.40} & 59.50 & 40.50 \\ 
  12 & France & 1.70 & 4.30 & 11.60 & \textbf{28.50} & 58.30 & 41.70 \\ 
  \hline
\end{tabular}
\caption{FIFA World Cup 2014 prediction via bivariate Poisson regression}
\end{table}

\begin{table}[h]
\centering
\begin{tabular}{llcccccc}
  \hline
 & Team & World Champion & Final &  Semi & Quarter & R16 & Prelim. Round \\ 
  \hline
1 & Spain & 20.70 & 31.40 & 43.30 & 57.60 & 88.00 & \textbf{12.00} \\ 
  2 & Brazil & 20.40 & 30.80 & \textbf{41.30} & 54.90 & 86.60 & 13.40 \\ 
  3 & Germany & \textbf{11.70} & 24.10 & 51.50 & 69.60 & 85.40 & 14.60 \\ 
  4 & Netherlands & 7.80 & 14.80 & \textbf{23.50} & 37.30 & 71.30 & 28.70 \\ 
  5 & Argentina & 7.70 & \textbf{16.80} & 38.00 & 67.60 & 93.70 & 6.30 \\ 
  6 & Portugal & 5.50 & 12.90 & 30.40 & 48.20 & 68.10 & \textbf{31.90} \\ 
  7 & Uruguay & 3.70 & 8.50 & 16.30 & 41.30 & \textbf{67.00} & 33.00 \\ 
  8 & England & 3.40 & 8.40 & 16.00 & 44.80 & 69.50 & \textbf{30.50} \\ 
  9 & Russia & 2.50 & 6.40 & 16.40 & 31.80 & 78.30 & \textbf{21.70} \\ 
  10 & Colombia & 1.80 & 4.50 & 9.60 & \textbf{27.70} & 59.90 & 40.10 \\ 
  11 & France & 1.80 & 4.30 & 11.80 & \textbf{28.60} & 58.80 & 41.20 \\ 
  12 & Italy & 1.70 & 4.20 & 9.20 & 25.80 & 48.80 & \textbf{51.20} \\ 
   \hline
\end{tabular}
\caption{FIFA World Cup 2014 prediction via diagonal inflated bivariate Poisson regression}
\end{table}

\newpage

\subsection{FIFA World Cup 2010 simulations}\label{app:10}
We present the tables of the forecast for the FIFA World Cup 2010 simulations using the bivariate Poisson regression model and the diagonal inflated bivariate Poisson regression model. Each model was simulated $100.000$ times.

\begin{table}[h!]
\centering
\begin{tabular}{rlcccccc}
  \hline
 & Team & World Champion & Final &  Semi & Quarter & R16 & Prelim. Round \\ 
  \hline
1 & Brazil & 24.60 & 33.20 & 43.90 & \textbf{60.90} & 77.30 & 22.70 \\ 
  2 & Netherlands & 15.10 & \textbf{24.40} & 35.90 & 65.20 & 89.10 & 10.90 \\ 
  3 & Spain & \textbf{12.00} & 24.60 & 35.10 & 52.70 & {93.20} & 6.80 \\ 
  4 & England & 6.60 & 13.50 & 30.20 & 50.10 & \textbf{77.40} & 22.60 \\ 
  5 & Italy & 5.20 & 11.60 & 20.60 & 47.90 & 85.00 & \textbf{15.00} \\ 
  6 & Portugal & 4.80 & 11.80 & 20.50 & 35.90 & \textbf{68.10} & 31.90 \\ 
  7 & France & 4.50 & 8.60 & 17.60 & 31.40 & 58.60 & \textbf{41.40} \\ 
  8 & Germany & 3.80 & 12.20 & \textbf{37.00} & 64.50 & 93.00 & 7.00 \\ 
  9 & Argentina & 3.60 & 10.40 & 28.30 & \textbf{52.60} & 87.40 & 12.60 \\ 
  10 & Uruguay & 3.20 & 6.70 & \textbf{15.20} & 29.20 & 56.40 & 43.60 \\ 
   \hline
\end{tabular}
\caption{World Cup 2010 prediction via bivariate Poisson regression}
\end{table}

\begin{table}[H]
\centering
\begin{tabular}{rlcccccc}
  \hline
 & Team & World Champion & Final &  Semi & Quarter & R16 & Prelim. Round \\ 
  \hline
1 & Brazil & 23.20 & 31.40 & 42.00 & \textbf{58.30} & 76.70 & 23.30 \\ 
  2 & Netherlands & 14.60 & \textbf{23.80} & 35.60 & 64.20 & 87.90 & 12.10 \\ 
  3 & Spain & \textbf{8.40} & 17.70 & 29.10 & 44.70 & 83.80 & 16.20 \\ 
  4 & England & 7.10 & 14.20 & 30.90 & 51.20 & \textbf{77.80} & 22.20 \\ 
  5 & Germany & 5.80 & 15.60 & \textbf{36.90} & 64.10 & 92.60 & 7.40 \\ 
  6 & Portugal & 5.30 & 12.50 & 21.90 & 38.20 & \textbf{68.00} & 32.00 \\ 
  7 & Italy & 5.30 & 11.70 & 21.30 & 47.50 & 84.70 & \textbf{15.30} \\ 
  8 & France & 4.60 & 8.60 & 17.40 & 31.40 & 58.50 & \textbf{41.50} \\ 
  9 & Argentina & 4.50 & 11.50 & 27.90 & \textbf{52.20} & 87.20 & 12.80 \\ 
  10 & Uruguay & 3.30 & 6.90 & \textbf{15.10} & 29.20 & 56.20 & 43.80 \\ 
    \hline
\end{tabular}
\caption{World Cup 2010 prediction via diagonal inflated bivariate Poisson regression}
\end{table}

\bibliographystyle{apalike}
\bibliography{bib}

\end{document}